\documentclass[reprint,10pt,NumberedRefs]{JASAnew}

\begin{document}

\title[Acoustic modes of rapidly rotating ellipsoids subject to centrifugal gravity]{Acoustic modes of rapidly rotating ellipsoids subject to centrifugal gravity}
\author{J\'er\'emie Vidal}
\email{jeremie.vidal@univ-grenoble-alpes.fr}
\author{David C\'ebron}
\affiliation{Universit\'e Grenoble Alpes, CNRS, ISTerre, 38000, Grenoble, France}


\date{\today} 
\DOInumber{10.1121/10.0005909}
\editorinitials{\href{https://doi.org/10.1121/10.0005909}{DOI:10.1121/10.0005909}}

\begin{abstract}
The acoustic modes of a rotating fluid-filled cavity can be used to determine the effective rotation rate of a fluid (since the resonant frequencies are modified by the flows). 
To be accurate, this method requires a prior knowledge of the acoustic modes in rotating fluids. 
Contrary to the Coriolis force, centrifugal gravity has received much less attention in the experimental context. 
Motivated by on-going experiments in rotating ellipsoids, we study how global rotation and buoyancy modify the acoustic modes of fluid-filled ellipsoids in isothermal (or isentropic) hydrostatic equilibrium. 
We go beyond the standard acoustic equation, which neglects solid-body rotation and gravity, by deriving an exact wave equation for the acoustic velocity.
We then solve the wave problem using a polynomial spectral method in ellipsoids, which is compared with finite-element solutions of the primitive fluid-dynamic equations. 
We show that the centrifugal acceleration has measurable effects on the acoustic frequencies when $M_\Omega \gtrsim 0.3$, where $M_\Omega$ is the rotational Mach number defined as the ratio of the sonic and rotational time scales. 
Such a regime can be reached with experiments rotating at a few tens of Hz, by replacing air with a highly compressible gas (e.g. SF$_6$ or C$_4$F$_8$).
\end{abstract}

\maketitle

\section{ Introduction}
\label{sec:Intro}
The acoustic resonant frequencies of fluid-filled cavities have several important applications in experimental physics.
The acoustic modes can, for instance, be employed to measure the thermodynamical properties of gases \cite{mehl1981precision,moldover1986gas}, to determine the Boltzmann constant in metrology \cite{moldover1988measurement,guianvarc2009acoustic,pitre2011determination}, or to passively determine the temperature in sulfur plasma lamps \cite{koulakis2018acoustic}. 
The acoustic modes are also known to be sensitive to the advection by background flows, which can slightly change the acoustic resonant frequencies (as evidenced by groundbreaking applications in astrophysics \cite{aerts2010asteroseismology}).
The acoustic modes have thus been recently used to remotely image the effective flow rotation rate (i.e. the differential rotation between the solid-body rotation and the flow) in rotating fluid-filled spheres  \cite{triana2014helioseismology} and spheroids \cite{su2020acoustic}. 
This imaging method promises future applications in experimental fluid mechanics, as it could be used for fluids that cannot be easily imaged with conventional velocimetry techniques (e.g. gases, which cannot be easily seeded by optical particles \cite{melling1997tracer}).

A theoretical knowledge of the diffusionless acoustic modes has proven important for the aforementioned applications, despite certain advances in computational acoustics (e.g. relying on finite-element computations \cite{su2020acoustic}). 
It is, for instance, still costly to numerically resolve the viscous (and thermal) boundary layers in three-dimensional models, which are thus often approximated using asymptotic theory \cite{moldover1986gas,berggren2018acoustic}.
Motivated by rapidly rotating fluid-filled experiments dedicated to planetary applications (e.g. the on-going ZoRo setup \cite{su2020acoustic} in spheroidal geometry), we aim to theoretically study the diffusionless acoustic modes in rapidly rotating ellipsoidal cavities subject to pressure or temperature variations. 
Acoustic modes with angular frequency $\omega$ are usually modeled using the equation
\begin{equation}
    - {\omega}^2 P_\omega = C_s^2 \, \nabla^2 P_\omega,
    \label{eq:helmholtz}
\end{equation}
where $P_\omega$ is the acoustic pressure and $C_s$ the adiabatic speed of sound.
Helmholtz Eq. (\ref{eq:helmholtz}) results from simple manipulations of the non-rotating primitive fluid equations upon a quiescent homogeneous medium, and analytical solutions can be obtained in spheroids \cite{chang1971natural,chang1972natural} and triaxial ellipsoids \cite{willatzen2004eigenmodes}.
The standard acoustic equation can also be simply extended when the background medium exhibits weak variations of pressure but strong density variations (i.e. when $|\nabla P_0| \ll |C_s^2 \nabla \rho_0|$, where $[P_0, \rho_0]$ are the background pressure and density). 
This gives the acoustic equation for waves upon isobaric states \cite{bergmann1946wave}
\begin{equation}
    -\omega^2 P_\omega = \rho_0 C_s^2 \, \nabla \boldsymbol{\cdot} \left ( \rho_0^{-1} \nabla P_\omega \right ),
    \label{eq:waveeqPierce}
\end{equation}
which is used in underwater acoustics or in the presence of strong density gradient of thermal origin\cite{karlsen2016acoustic,koulakis2018acoustic,koulakis2021convective}.
Equations (\ref{eq:helmholtz}) and (\ref{eq:waveeqPierce}) are, however, not valid when the fluid has substantial pressure variations in an ambient gravity field, or in the presence of global rotation.
A scalar acoustic equation for rotating stratified fluids has been obtained under the beta-plane approximation \cite{desanto1979derivation}, but it does not rigorously account for global rotation in three-dimensional inhomogeneous media (e.g. as encountered in the global fluid envelopes of rapidly rotating planets or stars). 
Rotational effects are thus usually modeled using asymptotic theory\cite{backus1961rotational} as small perturbations with respect to the acoustic modes upon non-rotating inhomogeneous media \cite{bergmann1946wave,lignieres2009asymptotic}, but the asymptotic approach unfortunately becomes inaccurate for rapidly rotating fluids (e.g. for rotating stars \cite{reese2006acoustic}, and even for uniform-density fluids \cite{vidal2020compressible}). 
A non-perturbative description of global rotation in (strongly) inhomogeneous media is thus desirable for rapidly rotating experiments in ellipsoidal geometries (but also in cylinders \cite{morton1972waves,miles1981waves,dodgson1988some}).

Going beyond the scalar acoustic equation is quite unconventional in acoustic modeling, but the usefulness of employing a vector wave equation has already been recognized in several contexts \cite{lynden1967stability,ross1986note,komatitsch2002spectral}. 
Therefore, we derive in this work an exact vector wave equation that includes a rigorous treatment of rotation and buoyancy for fluids in isentropic and isothermal equilibrium. 
Such equilibrium states are indeed relevant for experiments \cite{menaut2019experimental,su2020acoustic}.
The mathematical formulation extends our previous investigations of the acoustic modes in uniform-density fluids\cite{vidal2020acoustic} and in a planetary context\cite{vidal2020acoustic}, to include centrifugal gravity and isothermal reference states. 
We then solve the acoustic problem in ellipsoidal geometries, using a spectral description of the velocity field. 
The paper is organized as follows.
We describe the acoustic problem in \S\ref{sec:model}, then presents results in \S\ref{sec:results}. 
We discuss our results in light of experimental applications in \S\ref{sec:discussion}, and we end the paper in \S\ref{sec:conclusion}.

\section{Formulation of the problem}
\label{sec:model}
 \subsection{Primitive fluid-dynamic equations}
We consider a compressible Newtonian fluid of density $\rho$, pressure $P$ and temperature $T$, enclosed within a rigid arbitrary ellipsoidal cavity of semi-axes $[a,b,c]$ and volume $V=4\pi abc/3$.
The cavity is subject to the constant (local) Earth's gravity $\boldsymbol{g}_E = - g_E \boldsymbol{1}_z$ (where $\boldsymbol{1}_z$ is the unit vertical vector), and spins at the steady angular frequency $\boldsymbol{\Omega}=\Omega \, \boldsymbol{1}_z$ with respect to an inertial frame.  
To account for global rotation, we work in the frame rotating at $\boldsymbol{\Omega}$, where the ellipsoidal boundary $\partial V$ is steady and given by $(x/a)^2 + (y/b)^2 + (z/c)^2=1$ in the Cartesian coordinates $(x,y,z)$. 
Global rotation thus generates in the rotating frame the additional centrifugal acceleration  $\boldsymbol{g}_c = - \boldsymbol{\Omega} \times (\boldsymbol{\Omega} \times \boldsymbol{r})$, where $\boldsymbol{r}=(x,y,z)^\top$ is the position vector. 
We model sound below as small-amplitude time-dependent perturbations for the velocity $\boldsymbol{u}_1$, the density $\rho_1$, pressure $P_1$ and entropy $S_1$, upon a spatially inhomogeneous reference state $[\rho_0, P_0, S_0]$. 
The reference state is assumed to be in hydrostatic equilibrium
\begin{subequations}
\label{eq:hydrostatic}
\begin{equation}
    \nabla P_0 = \rho_0 \boldsymbol{g}, \quad \boldsymbol{g} = \boldsymbol{g}_E + \boldsymbol{g}_c,
    \tag{\theequation a,b}
\end{equation}
\end{subequations}
where $\boldsymbol{g}$ is the effective gravity. 
The other relation among the ambient variables is the equation of state (EoS) \cite{pierce1990wave}
\begin{equation}
    \nabla P_0 = C_s^2 \nabla \rho_0 + \left (  \frac{\partial P}{\partial S} \right )_\rho \nabla S_0,
    \label{eq:backgroundS0}
\end{equation}
where $S_0$ is the background entropy, and $C_s = \sqrt{(\partial P/\partial \rho)_S}$ is the adiabatic speed of sound that can be inhomogeneous in space. 

In the diffusionless theory (i.e. without attenuation due to viscosity and thermal diffusion), the small-amplitude perturbations are given in the rotating frame by the linearized fluid-dynamic equations \cite{pierce1990wave}
\begin{subequations}
\allowdisplaybreaks
\label{eq:primitiveeqn}
\begin{align}
\rho_0 \left ( \partial_t \boldsymbol{u}_1 + 2 \boldsymbol{\Omega} \times \boldsymbol{u}_1 \right ) &= - \nabla P_1 + \rho_1 \boldsymbol{g}, \\
  \partial_t P_1 + \boldsymbol{u} \boldsymbol{\cdot} \nabla P_0 &= - \rho_0 C_s^2 \, \nabla \boldsymbol{\cdot} \boldsymbol{u}_1, \\
 \partial_t \rho_1 + \nabla \boldsymbol{\cdot} (\rho_0 \boldsymbol{u}_1) &= 0,
\end{align}
\end{subequations}
where we have included the Coriolis term $2 \rho_0 \boldsymbol{\Omega} \times \boldsymbol{u}_1$ and the buoyancy force $\rho_1 \boldsymbol{g}$ in the momentum equation.
Note that Eq. (\ref{eq:primitiveeqn}b) is obtained from the isentropic equation of continuity $\partial_t S_1 + \boldsymbol{u}_1 \boldsymbol{\cdot} \nabla S_0 = 0$ for diffusionless perturbations. 
Additionally, the perturbations must also satisfy the EoS
\begin{equation}
    \partial_t P_1 = C_s^2 \partial_t \rho_1 +  \boldsymbol{u}_1 \boldsymbol{\cdot} \left [ C_s^2 \nabla \rho_0 - \nabla P_0 \right ],
\label{eq:eosdPdR}
\end{equation}
which is obtained by combining Eqs. (\ref{eq:backgroundS0}) and (\ref{eq:primitiveeqn}). 
The equations are also supplemented with appropriate boundary conditions (BC) on $\partial V$.
The boundary is assumed to be rigid (considering an infinite acoustic impedance at the wall).
This assumption is realistic for most experimental conditions (e.g. with a gas-filled metallic cavity\cite{su2020acoustic}).  
The fluid velocity thus obeys the no-penetration BC $\left . \boldsymbol{u}_1 \boldsymbol{\cdot} \boldsymbol{n} \right |_{\partial V} = 0$, where $\boldsymbol{n}=(x/a^2,y/b^2,z/c^2)^\top$ is the (non-unit) normal vector on the boundary. 
For a non-rotating fluid without buoyancy, this BC is equivalent to the sound hard BC $ \left . \nabla P_1 \boldsymbol{\cdot} \boldsymbol{n}\right |_{\partial V} = 0$ for the acoustic pressure.
However, the no-penetration BC gives a more complicated BC for the pressure in the presence of rotation and buoyancy (which is implicitly obtained from the continuity of the normal component of the momentum equation \cite{poinsot1992boundary}).
The BC for the density can finally obtained from the normal component of EoS (\ref{eq:eosdPdR}) if required. 

    \subsection{Wave equation for the acoustic modes}
The diffusionless acoustic modes are sought as harmonic perturbations $[\boldsymbol{u}_1, P_1, \rho_1] (\boldsymbol{r},t) = [\boldsymbol{u}_\omega, P_\omega, \rho_\omega] (\boldsymbol{r}) \exp(\lambda t)$ with $\lambda = \mathrm{i} \omega$, where $\omega$ is the angular frequency and $[\boldsymbol{u}_\omega, P_\omega, \rho_\omega] (\boldsymbol{r})$ represent the spatial structure of the modes. 
The identification of a master wave equation often allows a better understanding of the modes' properties, which are usually difficult to uncover from the primitive fluid-dynamic equations. 
As shown in Appendix \ref{appendix:waveeq}, it is possible to extend scalar acoustic Eqs. (\ref{eq:helmholtz}) and (\ref{eq:waveeqPierce}) to account for non-rotating inhomogenous reference states stratified in density under an external gravity field. 
The Coriolis force has however significant effects on the lowest-frequency acoustic modes \cite{vidal2020compressible,vidal2020acoustic}, which are important for experimental applications \cite{triana2014helioseismology,su2020acoustic}, and so should not be neglected (see below). 
Unfortunately, no simple scalar acoustic equation can be obtained in the presence of global rotation. 
To have a full account of rotational and buoyancy effects, we can fortunately combine the fluid-dynamic equations to obtain a rigorous wave equation for the velocity of the acoustic modes upon arbitrary hydrostatic reference states. 

To do so, we substitute Eqs. (\ref{eq:primitiveeqn}b) and (\ref{eq:primitiveeqn}c) into the time derivative of momentum Eq. (\ref{eq:primitiveeqn}a).
This gives
\begin{multline}
     \lambda^2 \boldsymbol{\zeta}_\omega = - 2 \lambda \boldsymbol{\Omega} \times \boldsymbol{\zeta}_\omega - (\nabla \boldsymbol{\cdot} \boldsymbol{\zeta}_\omega) \, \boldsymbol{g} + \nabla \left ( C_s^2 \, \nabla \boldsymbol{\cdot} \boldsymbol{\zeta}_\omega \right ) \\
     + \nabla \left ( \frac{1}{\rho_0} \boldsymbol{\zeta}_\omega \boldsymbol{\cdot} [\nabla P_0 - C_s^2 \, \nabla \rho_0] \right )
    \label{eq:waveeqU1}
\end{multline}
where the fluid momentum $\boldsymbol{\zeta}_\omega = \rho_0 \boldsymbol{u}_\omega$ is the dynamical unknown. 
The wave equation is here solely supplemented with the no-penetration BC $\left . \boldsymbol{\zeta}_\omega \boldsymbol{\cdot} \boldsymbol{n} \right |_{\partial V} = 0$, which is sufficient to solve the problem. 
Wave Eq. (\ref{eq:waveeqU1}) is an exact generalization of the scalar acoustic equations, which rigorously takes global rotation and buoyancy into account for arbitrary inhomogeneous reference states in hydrostatic equilibrium (i.e. in the presence of possibly large pressure and density gradients). 
Note that alternative vector equations have been obtained in astrophysics\cite{lynden1967stability} or seismology\cite{komatitsch2002spectral}, in which the fluid displacement vector is instead taken as unknown.

A few characteristics of Eq. (\ref{eq:waveeqU1}) are worth commenting on.  
The momentum can be expressed using Helmholtz decomposition \cite{vidal2020acoustic}
$\boldsymbol{\zeta}_\omega = \nabla \Phi_\omega + \nabla \times \boldsymbol{A}_\omega$, where $\Phi_\omega$ is the mass flux potential and $\boldsymbol{A}_\omega$ is the vector potential.
Acoustic studies generally seek the momentum in the form\cite{collas1987acoustic,pierce1990wave,karlsen2016acoustic,koulakis2018acoustic} $\boldsymbol{\zeta}_\omega = \nabla \Phi_\omega$, which is valid if the curl of all the terms in the right-hand side of Eq. (\ref{eq:waveeqU1}) vanishes.
This assumption is appropriate for non-rotating fluids with negligible pressure variations, such that we can set $\nabla P_0 = \boldsymbol{g} = \boldsymbol{0}$ in wave Eq. (\ref{eq:waveeqU1}) that then simply reduces to acoustic Eq. (\ref{eq:waveeqPierce}) for the mass flux potential (or the hydrodynamic pressure given by\cite{collas1987acoustic} $P_\omega = - \lambda \Phi_\omega$).
Here, this decomposition is however not appropriate because $\nabla \times \left [ 2 \lambda \boldsymbol{\Omega} \times \boldsymbol{\zeta}_\omega  + (\nabla \boldsymbol{\cdot} \boldsymbol{\zeta}_\omega) \, \boldsymbol{g} \right ] \neq \boldsymbol{0}$.
Indeed, the rotational component $\nabla \times \boldsymbol{A}_\omega$ must be retained in the Helmholtz decomposition of $\boldsymbol{\zeta}_\omega$ to obtain accurate solutions of the acoustic modes in rotating systems (even for uniform-density fluids \cite{vidal2020compressible}).
Therefore, we cannot reduce vector Eq. (\ref{eq:waveeqU1}) to a simple scalar equation in the presence of global rotation. 

    \subsection{Reference states}
We describe the inhomogeneous reference states that we consider in the following, and which are relevant for our experimental applications.
It is worth noting that the centrifugal forces do not modify here the boundary shape since the cavity is assumed to be perfectly rigid (contrary to stellar applications where free-surface flows are considered \cite{aerts2010asteroseismology}). 
Hence, we can solve the hydrostatic equilibrium assuming the problem geometry. 
To do so, we rewrite EoS (\ref{eq:backgroundS0}) in the form
\begin{equation}
    C_s^2 \nabla \rho_0 = \Gamma \, \nabla P_0,
    \label{eq:eospekeris}
\end{equation}
where $\Gamma$ is a thermodynamic parameter (possibly inhomogeneous in space).
EoS (\ref{eq:eosdPdR}) for the perturbations then reduces to $\partial_t P_1 = C_s^2 \partial_t + (\Gamma-1) \, \boldsymbol{u}_1 \boldsymbol{\cdot} \nabla P_0$ using EoS (\ref{eq:eospekeris}) for the ambient quantities, which clearly indicates that the linearized perturbations do not obey the standard relationship\cite{pierce1990wave} $\partial_t P_1= C_s^2 \, \partial_t \rho_1$ unless the reference state is characterized by $\Gamma=1$. 
This generic EoS for the reference state, initially introduced in the planetary context \cite{pekeris1972dynamics}, allows us to model two important experimental configurations. 

We can first model isentropic references states by setting $\Gamma=1$, which develop when the fluid is well-mixed by flow motions (e.g. by convective motions\cite{tilgner2011convection,menaut2019experimental}). 
Equilibrium (\ref{eq:eospekeris}) can also model isothermal ($\nabla T_0 = \boldsymbol{0}$) hydrostatic states for arbitrary fluids. 
Indeed, isothermal states are such that
\begin{equation}
    \nabla S_0 = \left (\frac{\partial S}{\partial P} \right )_T \nabla P_0,
\end{equation}
and Eq. (\ref{eq:backgroundS0}) can then be written in the form of Eq. (\ref{eq:eospekeris}) with
\begin{equation}
    \Gamma = 1-\left (\frac{\partial P}{\partial S} \right )_\rho \left (\frac{\partial S}{\partial P} \right )_T = 1-\frac{\alpha_P}{\alpha_S}.
    \label{eq:Gamma}
\end{equation}
We have used in the right-hand side of Eq. (\ref{eq:Gamma}) the Maxwell relations $\left . ( \partial P/ \partial S \right )_\rho = \rho^2 \left . ( \partial T/ \partial \rho \right )_S $ and
$\left . ( \partial S/ \partial P \right )_T = (1/\rho^2) \left . ( \partial \rho/ \partial T \right )_P $ to introduce
\begin{subequations}
\label{eq:thermalexpcoeffs}
\begin{equation}
    \alpha_S = - \frac{1}{\rho} \left ( \frac{\partial \rho}{\partial T} \right )_S, \quad \alpha_P = - \frac{1}{\rho} \left ( \frac{\partial \rho}{\partial T} \right )_P,
    \tag{\theequation a,b}
\end{equation}
\end{subequations}
where $\alpha_S$ and  $\alpha_P$ are respectively the isentropic\cite{ray1920isentropic,kouremenos1987isentropic} and isobaric coefficients of thermal expansion.
The thermodynamic parameter $\Gamma$ has not been directly tabulated in the literature, but it admits a simple expression.
As shown in Appendix \ref{appendix:Gamma}, we have indeed $\Gamma = \gamma$ for all fluids in isothermal equilibrium.
To summarize, we have thus 
\begin{equation}
    \Gamma = \begin{cases}
    1 & \text{for isentropic equilibrium}, \\
    \gamma & \text{for isothermal equilibrium}, \\
    \end{cases}
\end{equation}
in EoS (\ref{eq:eospekeris}) for all (diffusionless) fluids.

We can now simplify wave Eq. (\ref{eq:waveeqU1}) for isentropic and isothermal reference states, using Eq. (\ref{eq:hydrostatic}) and EoS (\ref{eq:eospekeris}).
We obtain after reduction
\begin{multline}
    \lambda^2 \boldsymbol{\zeta}_\omega + 2 \lambda \boldsymbol{\Omega} \times \boldsymbol{\zeta}_\omega  = \nabla \left [ C_s^2 \, \nabla \boldsymbol{\cdot} \boldsymbol{\zeta}_\omega  + (1-\Gamma) \, \boldsymbol{g} \boldsymbol{\cdot} \boldsymbol{\zeta}_\omega \right ] \\
    - (\nabla \boldsymbol{\cdot} \boldsymbol{\zeta}_\omega) \, \boldsymbol{g}.
    \label{eq:waveeqU2}
\end{multline}
Finally, it is worth noting that $\Gamma$ directly controls the density stratification of the fluid, which can affect the acoustic waves\cite{lignieres2009asymptotic}.
The strength of stratification is usually measured by the squared Brunt-V\"ais\"al\"a frequency \cite{komatitsch2002spectral}
\begin{equation}
    N_0^2 = \frac{1}{\rho_0} \left ( \nabla \rho_0 - \frac{\rho_0}{C_s^2} \boldsymbol{g} \right ) \boldsymbol{\cdot} \boldsymbol{g} = \frac{\Gamma-1}{C_s^2} \boldsymbol{g}^2
    \label{eq:BVN0}
\end{equation}
(with $\boldsymbol{g}^2 = \boldsymbol{g} \boldsymbol{\cdot} \boldsymbol{g}$), which quantifies the departure of the reference density field from an isentropic density profile. 
Neutral interiors $N_0^2=0$ are isentropic with $\Gamma=1$, whereas fluids in isothermal equilibrium with $\Gamma\neq1$ are stably stratified in density if $N_0^2 > 0$.
Since $\Gamma=\gamma$ in isothermal conditions (see in Appendix \ref{appendix:Gamma}), it shows that isothermal fluids are usually only slightly stably stratified in density (because $\gamma \leq 2$ for most fluids in standard experimental conditions).

    \subsection{Numerical modeling}
\begin{figure}
    \centering
    \includegraphics[width=0.42\textwidth]{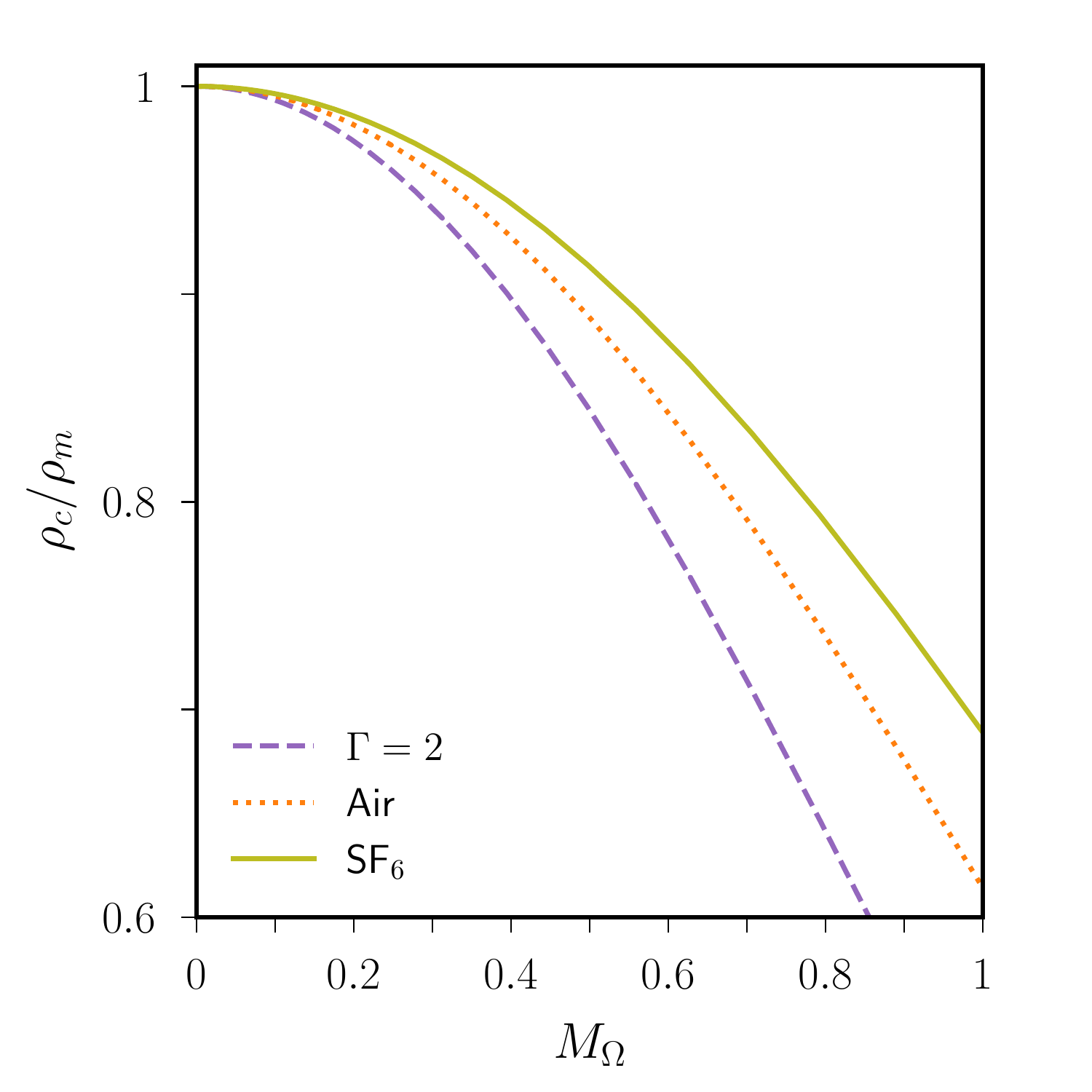}
    \caption{(Color online) Ratio $\rho_c/\rho_m$ given by conservation of mass (\ref{eq:rhocmass}), as a function of $M_\Omega$, for a rotating spheroid $a^*=b^*=1$ and $c^*=0.95$ in isothermal equilibrium with $M_G=0$.}
    \label{fig:rhoc}
\end{figure}

To solve wave Eq. (\ref{eq:waveeqU2}), we have to provide mathematical expressions for the isothermal and isentropic reference states. 
For mathematical simplicity, we assume below that $\gamma$ is constant in the isentropic and isothermal cases. 
For the isothermal case, we also assume that $C_s=C_0$ is spatially homogeneous (as encountered in most experimental conditions, e.g. for an ideal gas at constant temperature $T_0$).
The hydrostatic profiles for the isothermal equilibrium are then given by
\begin{equation}
[\rho_0, P_0] = \left [ \rho_c, P_c \right ] \exp \left ( \frac{s^2}{2\mathcal{H}_\Omega^2} \Gamma-\frac{z+c}{\mathcal{H}_G}  \Gamma  \right ),
\label{eq:densdav}
\end{equation}
where $\rho_c$ and $P_c = \rho_c C_0^2/{\Gamma}$ are respectively the density and pressure at $z=-c$ on the rotation axis, and with the cylindrical radius $s=(x^2+y^2)^{1/2}$. 
We have also introduced the gravity scale height $\mathcal{H}_G = {C_0^2}/g_E$, and the rotation scale height $\mathcal{H}_\Omega = {C_0}/{\Omega}$.
The value of $\rho_c$ is obtained by the conservation of mass
\begin{equation}
    \int_V \rho_0 \, \mathrm{d}V = (4/3) \pi abc \rho_m,
    \label{eq:rhocmass}
\end{equation}
where $\rho_m$ is the mean density of the fluid in the absence of rotation (and gravity). 
The ratio $\rho_c/\rho_m$, computed from Eq. (\ref{eq:rhocmass}), is illustrated in Fig. \ref{fig:rhoc} as a function of the strength of global rotation when $g_E = 0$. 
The density on the axis of rotation can be strongly weakened compared to the density at rest when the fluid is rapidly spinning. 

Contrary to the isothermal case, an isentropic equilibrium is characterized by an ambient inhomogeneous temperature $T_0$ defined by \cite{menaut2019experimental}
\begin{equation}
    \nabla T_0 = \frac{\alpha_P T_0}{C_P} \boldsymbol{g}.
    \label{eq:adiabaticT0}
\end{equation}
We assume below that the fluid obeys the ideal gas law $P_0=R_\star \rho_0 T_0$ in the isentropic case, where $R_\star=R/M$ is the specific gas constant (with the molar gas constant $R$ and the molar mass $M$ of the gas). 
We have thus $\alpha_P = 1/T_0$ and $C_P = R_\star\gamma/(\gamma-1)$, where $\gamma$ is the heat capacity ratio. 
The solution of Eq. (\ref{eq:adiabaticT0}) is then
\begin{equation}
    T_0 = T_c + \frac{\gamma-1}{R_\star \gamma} \left [ \frac{\Omega^2}{2} s^2 - g_E (z+c)\right ],
\end{equation}
where $T_c$ is the temperature at $s=0$ and $z=-c$.
The adiabatic speed of sound, defined by $C_s = (\gamma R_\star T_0)^{1/2}$ for an ideal gas, is thus given by the inhomogeneous profile
\begin{equation}
    C_s^2 = C_0^2 \left [ 1 + \frac{\gamma-1}{C_0^2} \left ( \frac{\Omega^2}{2}s^2 - g_E (z+c) \right ) \right ]
\end{equation}
for isentropic states (with the constant $C_0^2 = \gamma R_\star T_c$). 
Considering the dry air-filled ZoRo experiment\cite{su2020acoustic} rotating at $\Omega/(2\pi) = 50$~Hz, we obtain a temperature difference of $2$~K between the center and $s=0.2$~m. 

Having prescribed the reference state, Eq. (\ref{eq:waveeqU2}) is a quadratic eigenvalue problem of unknowns $[\lambda, \boldsymbol{\zeta}_\omega]$, which can be solved in ellipsoids as follows. 
We seek $\boldsymbol{\zeta}_\omega$ using the finite-dimensional expansion $\boldsymbol{\zeta}_\omega = \sum_j \Lambda_j \boldsymbol{e}_j$, where $\Lambda_j$ are complex-valued coefficients and $\boldsymbol{e}_j$ are vector elements that exactly satisfy the no-penetration BC $\left . \boldsymbol{e}_j \boldsymbol{\cdot} \boldsymbol{n} \right |_{\partial V} = 0$. 
The elements $\boldsymbol{e}_j$ are made of suitable combinations of Cartesian monomials $x^i y^jz^k$ of maximum degree $i+j+k\leq n$, which are sought using the Helmholtz decomposition \cite{vidal2020acoustic} $ \boldsymbol{e}_j = \nabla \Phi + \nabla \times \boldsymbol{\Psi}$ (where $\Phi$ is the velocity potential, and $\boldsymbol{\Psi}$ is the vector potential). 
The vortical part $\nabla \times \boldsymbol{\Psi}$, which is usually neglected in acoustics, must be retained in the presence of global rotation\cite{vidal2020compressible} or buoyancy since $\nabla \times (\lambda^2 \boldsymbol{\zeta}_\omega) \neq \boldsymbol{0}$ in the vector wave equation. 
We then substitute the polynomial expansion into wave Eq. (\ref{eq:waveeqU2}), and employ a projection (Galerkin) method to minimize the residuals with respect to every basis function $\boldsymbol{e}_i$ (using the volume-averaged inner product $\langle \boldsymbol{e}_i, \boldsymbol{e}_j \rangle = \int_V \boldsymbol{e}_i \boldsymbol{\cdot} \boldsymbol{e}_j \, \mathrm{d}V$). 
This gives a matrix quadratic eigenvalue problem
$\left [ \lambda^2 \boldsymbol{M} + \lambda \boldsymbol{C} + \boldsymbol{K} \right ] \boldsymbol{\Lambda} = \boldsymbol{0} $
for the eigenvalue $\lambda$ and the unknown vector $\boldsymbol{\Lambda} = (\Lambda_1, \Lambda_2,\dots)^\top$, which can be solved using standard numerical algorithms. 
We have implemented the aforementioned spectral method in our bespoke compressible numerical code \textsc{shine} \cite{vidal2020acoustic}, which has been extended to account for centrifugal gravity and isothermal reference states when $\Gamma \neq 1$. 
We truncate the polynomial expansion at maximum degree $n = 20$, which is sufficient to have a good frequency (and spatial) convergence \cite{vidal2020compressible} for the large-scale acoustic modes presented below. 
The above algorithm has already been thoroughly validated against theoretical and standard numerical computations for uniform-density\cite{vidal2020compressible} and inhomogeneous \cite{vidal2020acoustic} fluids, but neither in the presence of centrifugal effects nor when $\Gamma\neq 1$ in Eq. (\ref{eq:waveeqU2}). 
To validate the algorithm in such conditions, we have also employed standard finite-element computations performed with \textsc{comsol} (see the numerical details in Appendix \ref{appendix:comsol}).

\section{Results}
\label{sec:results}
We present below numerical computations of the resonant acoustic frequencies. 
To survey the parameter space, it is advantageous to non-dimensionalize the physical variables. 
We use the semi-axis $a$ as the length scale, the typical value $C_0$ of the speed of sound as the velocity scale, the sonic timescale $a/C_0$, the density $\rho_c$ at $s=0$ and $z=-c$ as the density scale, and $P_c$ as the pressure scale. 
Rotational and buoyancy effects are then controlled by the dimensionless numbers
\begin{subequations}
\label{eq:adimnumbers}
\begin{equation}
    M_\Omega = \frac{a \Omega}{C_0} = \frac{a}{\mathcal{H}_\Omega}, \quad M_G = \frac{\sqrt{a g_E}}{C_0} = \sqrt{\frac{a}{\mathcal{H}_G}},
    \tag{\theequation a,b}
\end{equation}
\end{subequations}
where $M_\Omega$ is the rotational Mach number (which compares the sonic time scale and the rotational time scale $\Omega^{-1}$), and $M_G$ is the gravitational Mach number (i.e. the ratio of the free-fall velocity $\sqrt{a g_E}$ and the speed of sound).
In dimensionless form, the Coriolis force scales as $M_\Omega$ while the centrifugal acceleration varies as $M_\Omega^2$, and the buoyancy force due to Earth's gravity evolves as $M_G^2$.
Typical experimental values give $M_\Omega \leq 1$ and $M_G \ll 1$. 
The dimensionless variables are written below with an asterisk ${}^*$ for clarity (to distinguish them from their dimensional counterparts).

Since the angular frequencies come in pairs $\pm \omega^*$ (by virtue of the symmetries of the problem), we only consider the solutions $\omega^* > 0$ in the following.
To compute the acoustic modes, an ellipsoidal geometry must be specified.
Solutions of acoustic Eq. (\ref{eq:helmholtz}) have angular frequencies that exhibit an azimuthal degeneracy in non-rotating spheres (i.e. the modes with different azimuthal wave numbers, but with the same meridional structure, have identical angular frequencies).
However, an ellipsoidal deformation is known to lift the azimuthal degeneracy of the acoustic frequencies, which allows a simple identification of the acoustic modes in the ellipsoid using their observed frequencies \cite{su2020acoustic}. 
Moreover, ellipsoidal geometries are also relevant for planetary applications\cite{le2015flows}. 
The effects of ellipsoidal deformation have already been thoroughly studied elsewhere \cite{vidal2020compressible,vidal2020acoustic} (albeit without centrifugal gravity), and so will not be further presented here. 
We consider below the ellipsoidal geometry of the ZoRo apparatus\cite{su2020acoustic}, a spheroid with dimensionless axes $a^*=b^*=1$ and $c^*=0.95$, and we have checked that similar conclusions are obtained in other geometries (not shown). 

\subsection{Effects of Earth's gravity}
\begin{figure}
    \centering
    \includegraphics[width=0.49\textwidth]{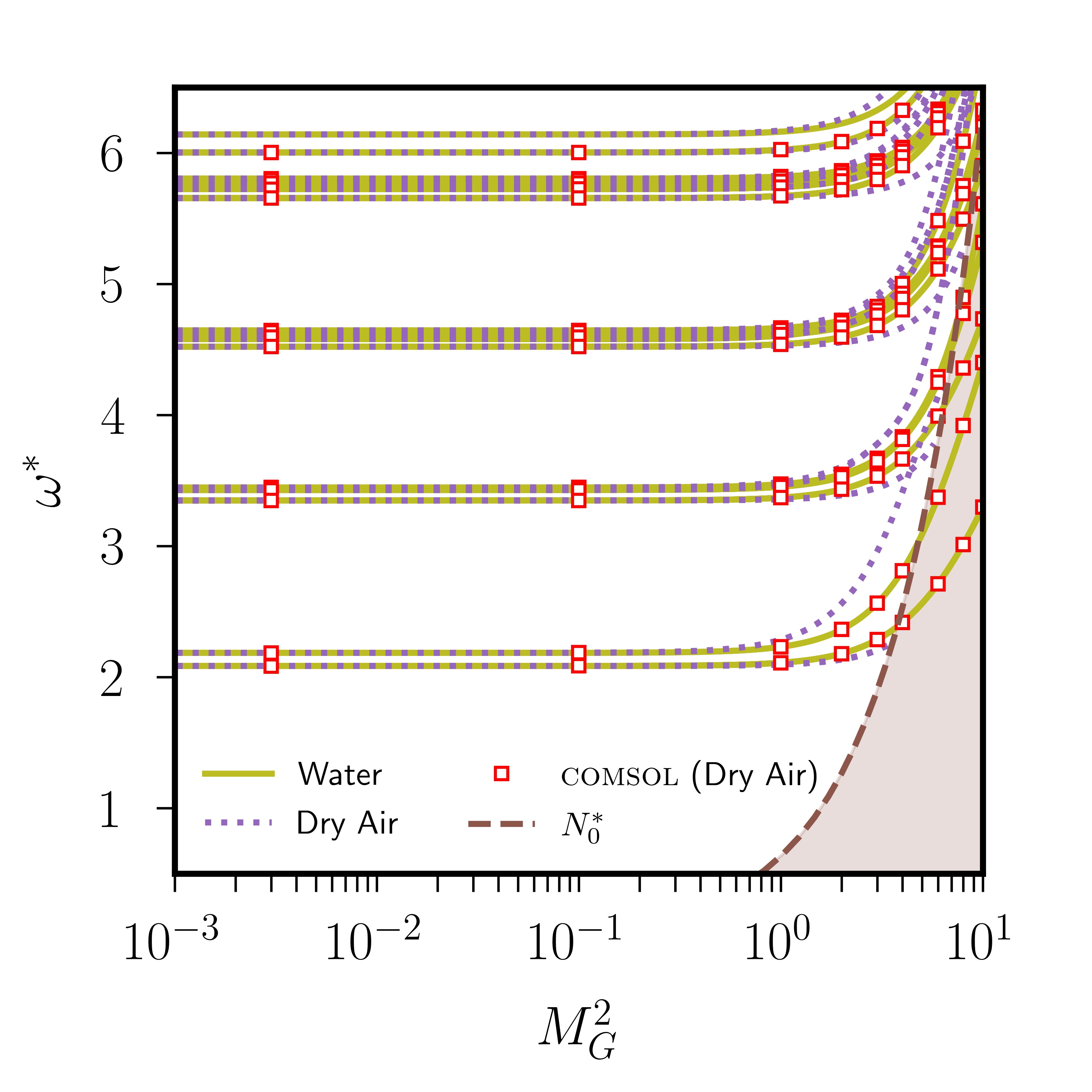}
    \caption{(Color online) Dimensionless angular frequency $\omega^*$ of the lowest-frequency acoustic modes, as a function of $M_G^2$ for isothermal reference states in a non-rotating spheroid with dimensionless polar axis $c^*=0.95$. 
    Red empty squares are the \textsc{comsol} solutions of Eq. (\ref{eq:waveeqP}) with $\boldsymbol{g}=\boldsymbol{g}_E$ and $\Gamma=1$, for azimuthal wave numbers $|m|\leq 8$. Brown area represents internal gravity modes, which exist when $\Gamma \neq 1$ in the range $|\omega^*|\leq N_0^*$ with $N_0^* = M_G^2 \sqrt{\Gamma-1}$.}
    \label{fig:nonrotating}
\end{figure}

Earth's gravity is usually neglected in the description of the non-rotating acoustic modes\cite{bergmann1946wave,pierce1990wave}, though it may modify the acoustic modes at rest for sufficient large values of $M_G$ (due to the presence of buoyancy terms in the wave equation). 
We explore in Fig. \ref{fig:nonrotating} how the angular frequencies of the non-rotating acoustic modes upon isothermal reference states vary with $M_G^2$ (for dry air with $\Gamma =1.4$ and water with $\Gamma \simeq 1$).
We have also shown the \textsc{comsol} solutions for $\Gamma=1$ obtained by solving Eq. (\ref{eq:waveeqP}) with $\boldsymbol{g}=\boldsymbol{g}_E$, which are in perfect agreement with the polynomial solutions. 
We find that the modes are almost unaffected by Earth's gravity for small values $M_G^2 \ll 1$ (as expected), but Earth's gravity is able to modify the acoustic frequencies when typically $M_G \gtrsim \mathcal{O}(1)$ for all the modes (the precise value is mode-dependent). 
The acoustic frequencies are also more affected for air than for liquid water (as expected since air is more compressible). 
Another interesting point in the figure is the presence of non-acoustic modes when $\Gamma \neq 1$, which have non-discrete frequencies in the spectrum (contrary to the acoustic modes). 
These modes are internal gravity modes that exist because the fluid is stably stratified in density, as evidenced here by the non-zero value of the Brunt-V\"ais\"al\"a frequency $N_0^* = M_G^2 \sqrt{\Gamma-1}$.
Since the spectrum of internal gravity modes is bounded\cite{friedlander1982internal} by $|\omega^*| \leq N_0^*$, they can have frequencies comparable to the acoustic ones when $M_G$ is sufficiently large. 

We have shown that Earth's gravity has measurable effects when $M_G \gtrsim \mathcal{O}(1)$, that is when $a g_E \gtrsim C_0^2$. 
Such an extreme regime is valid for atmospheric conditions where the length scale is very large, but it cannot be obtained in experimental conditions (even if a strongly compressible gas is used such as SF$_6$, see its properties in Table \ref{tab:thermogas}). 
Therefore, we discard Earth's gravity in the following and set $M_G = 0$ in all the computations. 

\subsection{Effects of centrifugal gravity}
\begin{figure}
    \centering
    \includegraphics[width=0.49\textwidth]{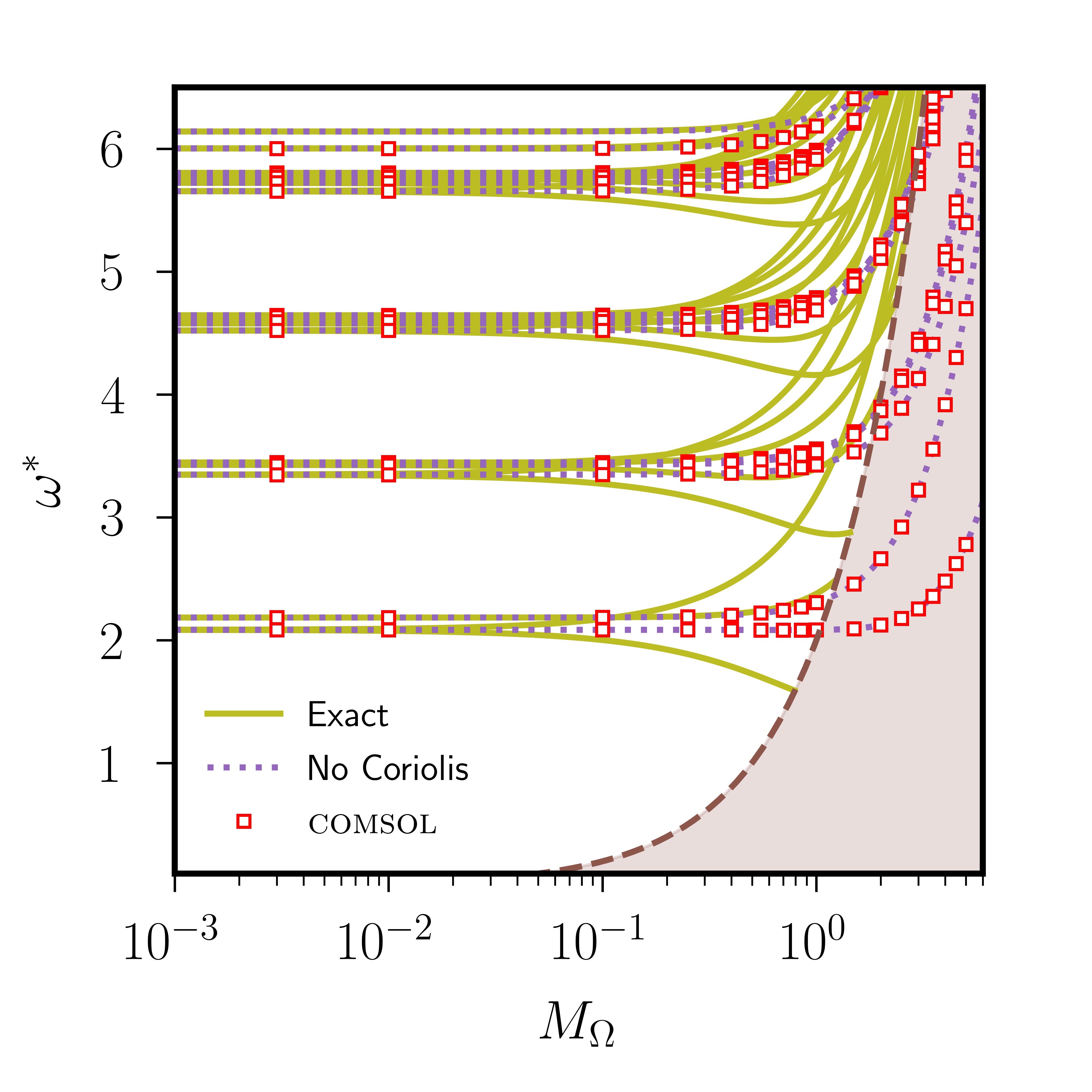}
    \caption{(Color online) Dimensionless angular frequency $\omega^*$ of acoustic modes in a rotating spheroid with dimensionless polar axis $c^*=0.95$, as a function of $M_\Omega$ for dry air ($\gamma=1.4$) in isentropic equilibrium. 
    Olive solid curves represent numerical solutions of exact Eq. (\ref{eq:waveeqU2}) without approximation, whereas dashed curves indicate solutions without the Coriolis force. 
    In the presence of the Coriolis force, some acoustic modes can hybridize with inertial modes that exist when \cite{vidal2020acoustic} $|\omega^*|\leq 2 M_\Omega$ (brown area).
    Red empty squares are \textsc{comsol} solutions of scalar Eq. (\ref{eq:waveeqP}) with $\boldsymbol{g}=\boldsymbol{g}_c$ and azimuthal wave numbers $|m| \leq 6$.}
    \label{fig:rotating1}
\end{figure}

Contrary to Earth's gravity, we cannot \emph{a priori} neglect the effects of centrifugal gravity on the acoustic modes in rapidly rotating configurations.
The centrifugal acceleration has indeed a maximum amplitude $\Omega^2 a$, which can be hundred times larger than Earth's gravity for rapid rotation (when $\Omega/(2\pi) \geq 11$~Hz for the ZoRo apparatus\cite{su2020acoustic} with $a = 0.2$~m).
Actually, the centrifugal and Coriolis effects have received much attention in the modeling of stellar oscillations. 
Stellar models usually neglect the Coriolis force in the wave equation whereas the buoyancy force (including centrifugal gravity) is retained, because the Coriolis force is expected to have a weak influence on the high-frequency acoustic modes that are used in astrophysics\cite{lignieres2009asymptotic}.

We assess the validity of this approximation in Fig. \ref{fig:rotating1} for dry air in isentropic equilibrium.  
We show the evolution of $\omega^*$ as a function of $M_\Omega$ for some acoustic modes obtained from vector Eq. (\ref{eq:waveeqU2}), either without approximation (i.e. with a rigorous description of the Coriolis and centrifugal effects) or artificially neglecting the Coriolis force.
We also show the \textsc{comsol} solutions of acoustic Eq. (\ref{eq:waveeqP}) with $\boldsymbol{g}=\boldsymbol{g}_c$, which are in excellent quantitative agreement with the solutions of the vector wave equation without the Coriolis force (as expected). 
The results clearly show that neglecting the Coriolis force is physically incorrect when $M_\Omega \geq \mathcal{O}(10^{-2})$, since the lowest-frequency modes of experimental interest\cite{triana2014helioseismology,su2020acoustic} are strongly impacted by the Coriolis force in this range \cite{vidal2020compressible}.
Note that some acoustic modes can also hybridize, when $M_\Omega \geq 1$, with the inertial modes\cite{greenspan1968theory}.
The latter are fluid modes sustained by the Coriolis force, which are characterized by the frequency spectrum  $|\omega^*| \leq 2 M_\Omega$ in compressible fluids \cite{vidal2020acoustic}.
The inertial modes can thus interact with some low-frequency acoustic modes in that range, which would make the identification of the proper acoustic modes more difficult for experimental applications. 

\begin{figure}
    \centering
    \includegraphics[width=0.49\textwidth]{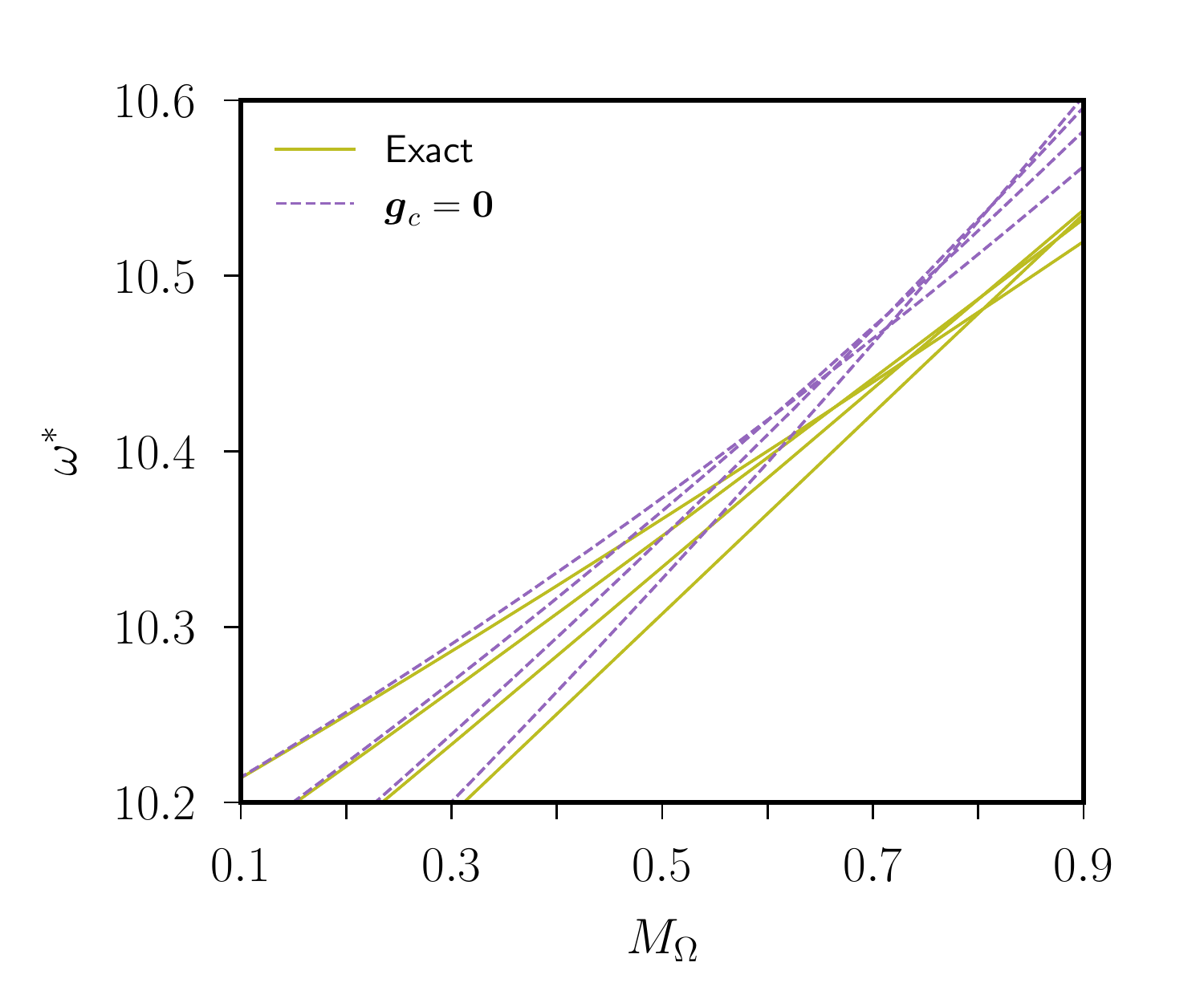}
    \caption{(Color online) Effects of centrifugal gravity on the dimensionless angular frequency $\omega^*$ of a few high-frequency acoustic branches in a rotating spheroid with dimensionless polar axis $c^*=0.95$, as a function of $M_\Omega$ for dry air ($\gamma=1.4$) in isothermal equilibrium.  
    Olive solid curves represent solutions of exact Eq. (\ref{eq:waveeqU2}) without approximation, whereas dashed curves indicate solutions without the centrifugal force for the perturbation (i.e. setting $\boldsymbol{g}_c=\boldsymbol{0}$ in the equation).}
    \label{fig:rotating2}
\end{figure}

The relative importance of centrifugal gravity must be finally addressed, since it could have weaker effects than reported in stellar applications (since centrifugal acceleration does not deform here the rigid cavity).
In our case, it could indeed be argued that centrifugal gravity would only slightly modify the acoustic resonant frequencies, since the buoyancy force $\rho_1 \boldsymbol{g}_c$ in the momentum equation has the typical amplitude $M_\Omega^2$ (which is usually smaller than the amplitude of the Coriolis force $M_\Omega$). 
As previously reported in astrophysics\cite{reese2006acoustic}, we find that the centrifugal effects are more pronounced for the high-frequency acoustic modes (in particular because they have smaller-scale structures). 
As shown in Fig. \ref{fig:rotating2}, centrifugal gravity can have measurable effects onto the acoustic frequencies of the high-frequency modes as soon as $M_\Omega \gtrsim 0.3$.
We have illustrated in the figure a few high-frequency branches with $\omega^* \simeq 10$, which approximately correspond to the highest-frequency modes that can be currently detected in rotation using the acoustic device of the ZoRo apparatus \cite{su2020acoustic}. 
The frequencies of the illustrated modes is erroneously shifted towards higher values if centrifugal gravity is disregarded, and we obtain crossings of acoustic branches that occur at wrong values of $M_\Omega$. 
Erroneous mode crossings can also occur for lower-frequency modes (e.g. in Fig. \ref{fig:diffusion} below).
Mode crossing is a phenomenon that considerably complicates the identification of the modes in the experimental spectrum \cite{su2020acoustic}, and will also raise mathematical issues for the interpretation of the observed resonant frequencies in terms of flow structures for modal acoustic velocimetry \cite{vidal2020compressible}. 
A theoretical model accounting for centrifugal gravity is thus appropriate to obtain accurate predictions for the acoustic frequencies, and also to precisely locate the possible mode crossings in the parameter space. 

\section{Discussion}
\label{sec:discussion}
\subsection{Effects of attenuation}
\begin{table}
    \centering
    \caption{Physical properties of fluids usually used in laboratory experiments, at pressure $10^5$ Pa and ambient temperature $T_0=20 \,^\circ$C. 
    Heat capacity ratio $\gamma$, specific gas constant $R_\star$, mean density at rest $\rho_m$, typical adiabatic speed of sound $C_0$, dynamic shear viscosity $\mu$, thermal conductivity $k$, specific heat at constant pressure $C_P$, and dimensionless Prandtl number $Pr=\nu/\kappa = C_p \mu/k$ where $\nu=\mu/\rho_m$ is the kinematic viscosity and $\kappa=k/(\rho_m C_P)$ is the thermal diffusivity. 
    Properties of SF$_6$ are taken from the library CoolProp\cite{bell2014pure}.}
    \begin{tabular}{cccc}
    \hline
    \hline 
     Parameter & Water & Dry Air & SF$_6$ \\
    \hline
    $\gamma-1$ & $7 \times 10^{-3}$ & $0.4$ & $0.1$ \\
    $ R_\star$ (J.kg$^{-1}$.K$^{-1}$) & $461.5$ & $287.0$ & $56.93$ \\
    $\rho_m$ (kg.m$^{-3}$) & $10^3$ & $1.204$ & $6.07$ \\
    $C_0$ (m.s$^{-1}$) & $1481$ & $343.2$ & $133.9$ \\
    $\mu$ (Pa.s) & $8.9 \times 10^{-4}$ & $1.8 \times 10^{-5}$ & $1.5 \times 10^{-5}$ \\
    $k$ (W.K$^{-1}$.m$^{-1}$) & $0.6$ & $2.58\times 10^{-2}$ & $1.26 \times 10^{-2}$ \\
    $C_P$ (J.kg$^{-1}$.K$^{-1}$) & $4.18 \times 10^3$ & $1005.42$ & $660.99$ \\
    $\alpha_P$ (K$^{-1}$) & $2.07 \times 10^{-4}$ & $3.42 \times 10^{-3}$ & $3.55 \times 10^{-3}$ \\
   $Pr$ & $6.2$ & $0.7$ & $0.8$ \\
   \hline
   \hline
    \end{tabular}
    \label{tab:thermogas}
\end{table}

We have shown that centrifugal gravity has non-negligible effects on the diffusionless acoustic frequencies when $M_\Omega \gtrsim 0.3$. 
The validity of the diffusionless theory may however be questioned for experimental applications, where fluid viscosity and thermal diffusion are present (see some typical values in Table \ref{tab:thermogas}). 
The predicted centrifugal effects might indeed be buried in the experimental noise (e.g. if diffusive effects were stronger than centrifugal effects).
Hence, it is important to explore the effects of diffusion on the acoustic modes.

\begin{figure}
    \centering
    \includegraphics[width=0.49\textwidth]{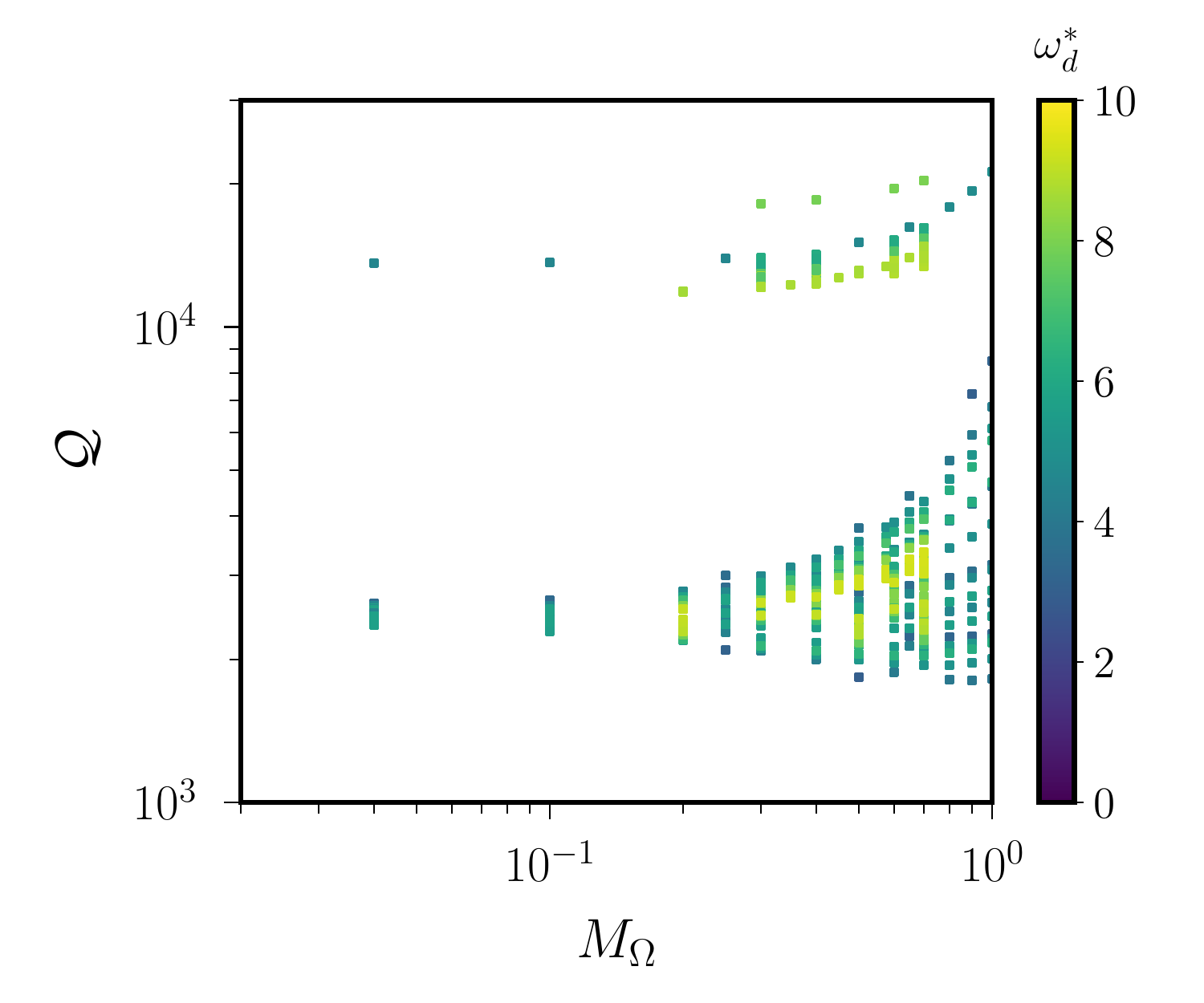}
    \caption{(Color online) Quality factor $\mathcal{Q}$, as a function of $M_\Omega$, for the diffusive acoustics modes with azimuthal wave numbers $|m| \leq 6$ in a rotating spheroid with dimensionless polar axis $c^*=0.95$. 
    \textsc{comsol} solutions are obtained from the diffusive fluid-dynamic equations for dry air in isothermal equilibrium at ambient temperature $T_0=20 \,^\circ$C (see in Table \ref{tab:thermogas}). 
    Color bar shows the diffusive angular frequency $\omega_d^*$.}
    \label{fig:quality}
\end{figure}

To address this point, we have computed acoustic modes from the primitive fluid-dynamic equations in the presence of viscosity and thermal diffusion, for dry air at ambient temperature $T_0=20^\circ$~C using \textsc{comsol} (see details in Appendix \ref{appendix:comsol}). 
We have considered acoustic modes with large-scale azimuthal wave numbers $|m|\leq 6$ and angular frequencies $|\omega^*| \leq 10$, which correspond to the modes that could be identified with the acoustic apparatus of the ZoRo experiment \cite{su2020acoustic}. 
The eigenvalue of a diffusive mode becomes $\lambda^* = \sigma^* + \mathrm{i} \omega_d^*$, where $\sigma^* < 0$ is the damping rate of the mode (due to viscous effects and thermal diffusion) and $\omega_d^*$ is the diffusive angular frequency. 
We illustrate in Fig. \ref{fig:quality} the evolution of the quality factor 
\begin{equation}
   \mathcal{Q} = |\omega_d^*|/|\sigma^*|
\end{equation}
as a function of $M_\Omega$. 
Very large quality factors $\mathcal{Q} \geq 10^{3}$ are found when $M_\Omega \leq 1$, showing that these modes can be measured in experimental conditions. 

\begin{figure*}
    \centering
    \begin{tabular}{cc}
    \includegraphics[width=0.49\textwidth]{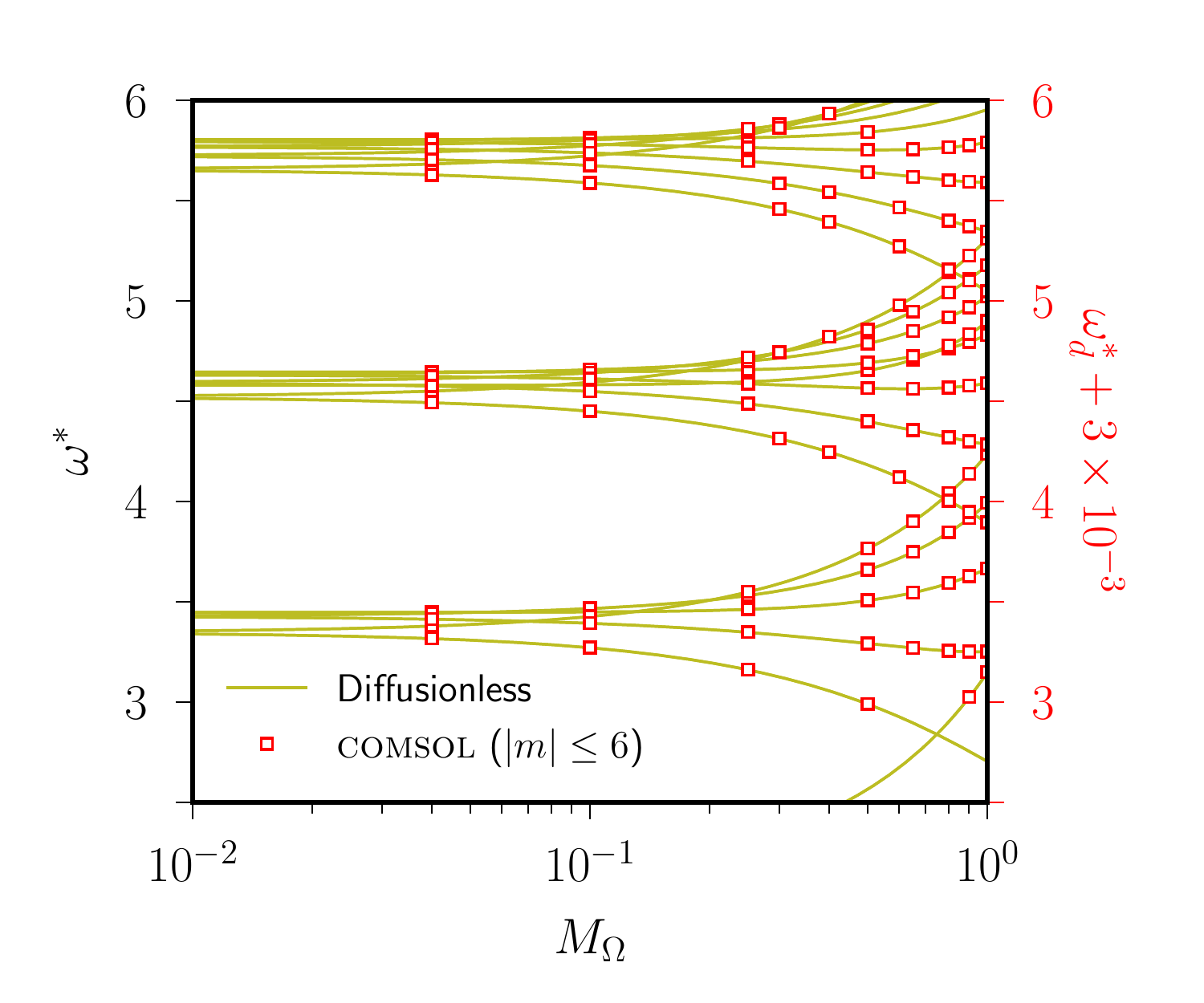} &
    \includegraphics[width=0.49\textwidth]{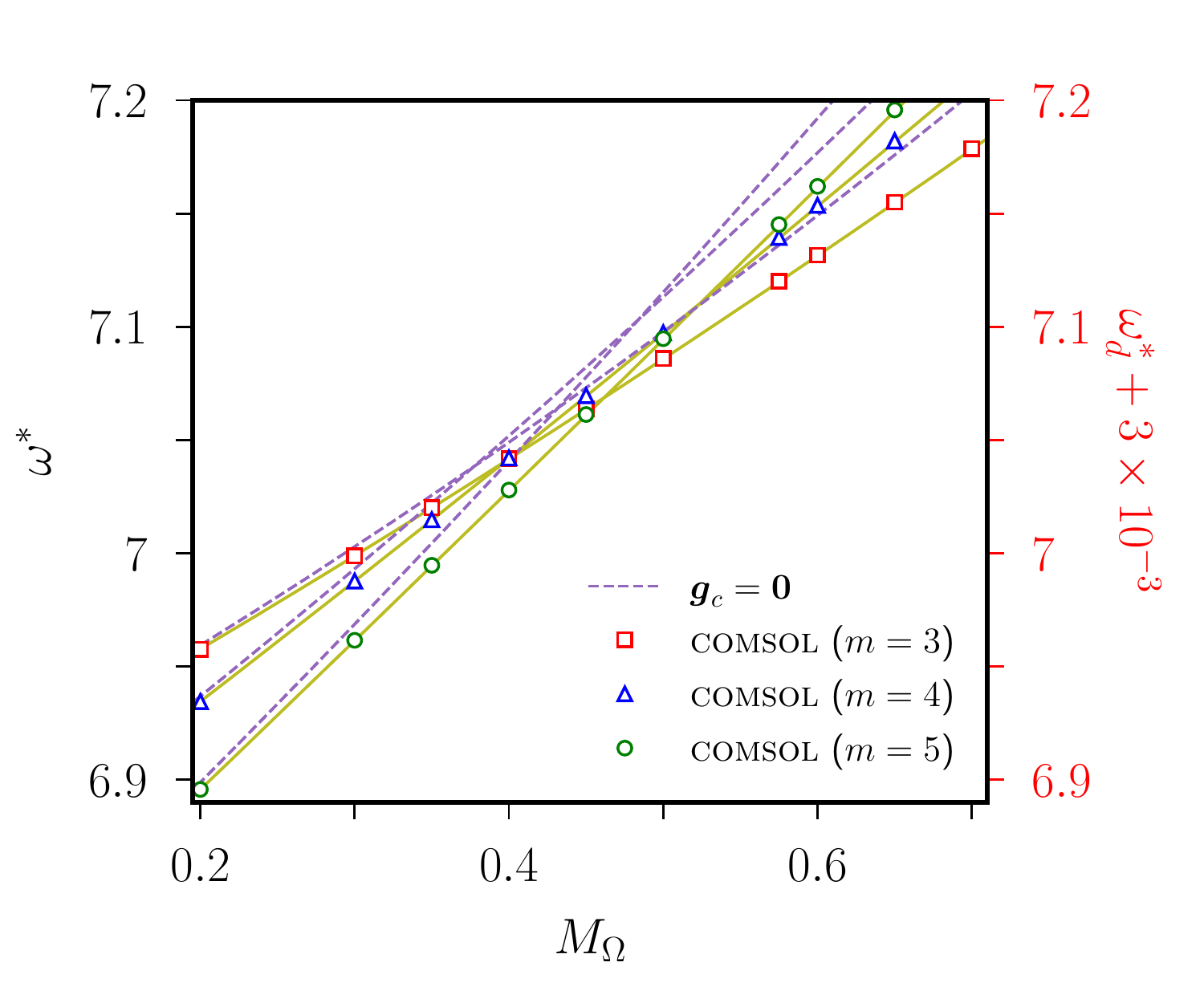} \\
    \end{tabular}
    \caption{(Color online) Comparison between diffusionless (left vertical axis, colored curves) and diffusive (right vertical axis, red empty squares) angular frequencies of some acoustic modes in a rotating spheroid with dimensionless polar axis $c^*=0.95$, as a function of $M_\Omega$ for isothermal states with $\Gamma=1.4$ (dry air). 
    Olive solid curves are diffusionless solutions of vector wave Eq. (\ref{eq:waveeqU2}).
    Empty symbols are \textsc{comsol} solutions with azimuthal wave numbers $|m| \leq 6$ of the exact diffusive fluid-dynamic equations for dry air at ambient temperature $T_0=20 \,^\circ$C (see parameters in Table \ref{tab:thermogas}). 
    For graphical purposes, the diffusive frequencies $\omega_d^*$ have been shifted by the amount $3\times 10^{-3}$ on the right vertical axis (to be superimposed with the diffusionless frequencies).
    In the right panel, dashed colored curves indicate diffusionless solutions without the centrifugal force for the perturbation (as in Fig. \ref{fig:rotating2}).}
    \label{fig:diffusion}
\end{figure*}

In addition to the damping $\sigma^*$, attenuation is also known to slightly reduce the acoustic resonant frequencies. 
Diffusion is responsible for small shifts in frequency $0 \leq \delta_\omega^* \ll \mathcal{O}(1)$ of the diffusionless frequencies $\omega^*$ in the weak attenuation regime, such that the actual diffusive acoustic frequency is $\omega_d^* \approx \omega^* - \delta_\omega^*$ at leading asymptotic order \cite{moldover1986gas,su2020acoustic}. 
The comparison between the diffusionless theory and the diffusive solutions of the exact diffusive equations is illustrated in Fig. \ref{fig:diffusion}. 
To ease the comparison between our diffusionless and diffusive computations, we have not directly shown in the figure the diffusive frequencies $\omega_d^*$ but instead $\omega_d^* + \delta_\omega^*$, where $\delta_\omega^* \simeq 3 \times 10^{-3}$ is the typical shift of frequency due to attenuation. 
An excellent quantitative agreement is then observed between the diffusionless frequencies and diffusive frequencies, showing that attenuation actually only very weakly depends on $M_\Omega$ for the modes with $|\omega^*| \leq 10$.
Moreover, as further illustrated in the right panel for a few high-frequency modes with $|\omega^*| \simeq 7$, centrifugal effects do persist in the presence of attenuation when $M_\Omega \gtrsim 0.2-0.3$. 
Therefore, the effects of centrifugal gravity must be considered for rapidly rotating experiments. 

\subsection{Parameter regimes for experiments}
It is worth estimating if the regime $M_\Omega \gtrsim 0.3$ where centrifugal effects are significant is relevant for on-going laboratory experiments dedicated to planetary applications. 
Measuring the resonant acoustic frequencies can indeed be used to reconstruct the effective rotation profile of the fluid \cite{triana2014helioseismology}, which requires prior accurate predictions of the resonant frequencies in the presence of solid-body rotation. 
Existing rotating experiments in ellipsoidal geometries are usually filled with water\cite{lemasquerier2017libration} or dry air\cite{su2020acoustic}. The  typical parameters for water-filled experiments are $a \lesssim 0.2$~m and $\Omega/(2\pi)\leq 4$~Hz, such that $M_\Omega \ll 10^{-2}$.
The effects of centrifugal gravity on the acoustic frequencies is thus entirely negligible for such experiments (as well as the Coriolis effects\cite{vidal2020compressible}).
The gas-filled acoustic experiment ZoRo, with $a=0.2$~m and $\Omega/(2\pi)\leq 70$~Hz, can reach larger values $M_\Omega \leq 0.25$ when filled with dry air at ambient temperature, but this is still insufficient to be sensitive to the centrifugal effects. 

However, it is currently under consideration to fill the ZoRo experiment with gases having much smaller speeds of sound than air, such as SF$_6$ or C$_4$F$_8$ with respectively\cite{bell2014pure} $C_0 \simeq 133.9$~m.s$^{-1}$ and $C_0 \simeq 110.3$ m.s$^{-1}$ at the usual experimental conditions of pressure $P_0 = 10^5$~Pa and temperature $T_0=20^\circ$~C.
Much larger values $M_\Omega \leq 0.7$ could thus be reached with the ZoRo experiment rotating at high rotation rates and filled with such gases. 
Therefore, it will be imperative to account for centrifugal gravity in forthcoming experiments that will consider gases more compressible than air in the reachable rapidly rotating regime $ M_\Omega \gtrsim 0.3$. 
Using other gases would also allow us to explore different regimes for the flow dynamics. 
For instance, SF$_6$ is characterized by a much smaller value of the kinematic viscosity $\nu=\mu/\rho_m \simeq 2.5 \times 10^{-6}$ (compared to $\nu \simeq 1.5 \times 10^{-5}$ for dry air).
Changing $\nu$ will indeed give a different force balance between viscous diffusion and rotation, as measured by the Ekman number $E=\nu/(a^2 \Omega)$ that plays a central role in the theory of rotating fluids\cite{greenspan1968theory}.
Since it is very small in planetary bodies (typically $E\ll10^{-10}$), reducing $\nu$ will significantly enhance the planetary relevance of the ZoRo experiment.
SF$_6$ also has a smaller thermal diffusivity $\kappa=k/(\rho_m C_P)$ than air (see in Table \ref{tab:thermogas}), which would allow us to change the value of Prandtl number $Pr = \nu/\kappa$ (which strongly impacts the outcome of buoyancy-driven flows, e.g. for thermal convection\cite{kaplan2017subcritical}).
Therefore, using different gases is highly desirable for planetary modeling.

\begin{figure}
    \centering
    \includegraphics[width=0.49\textwidth]{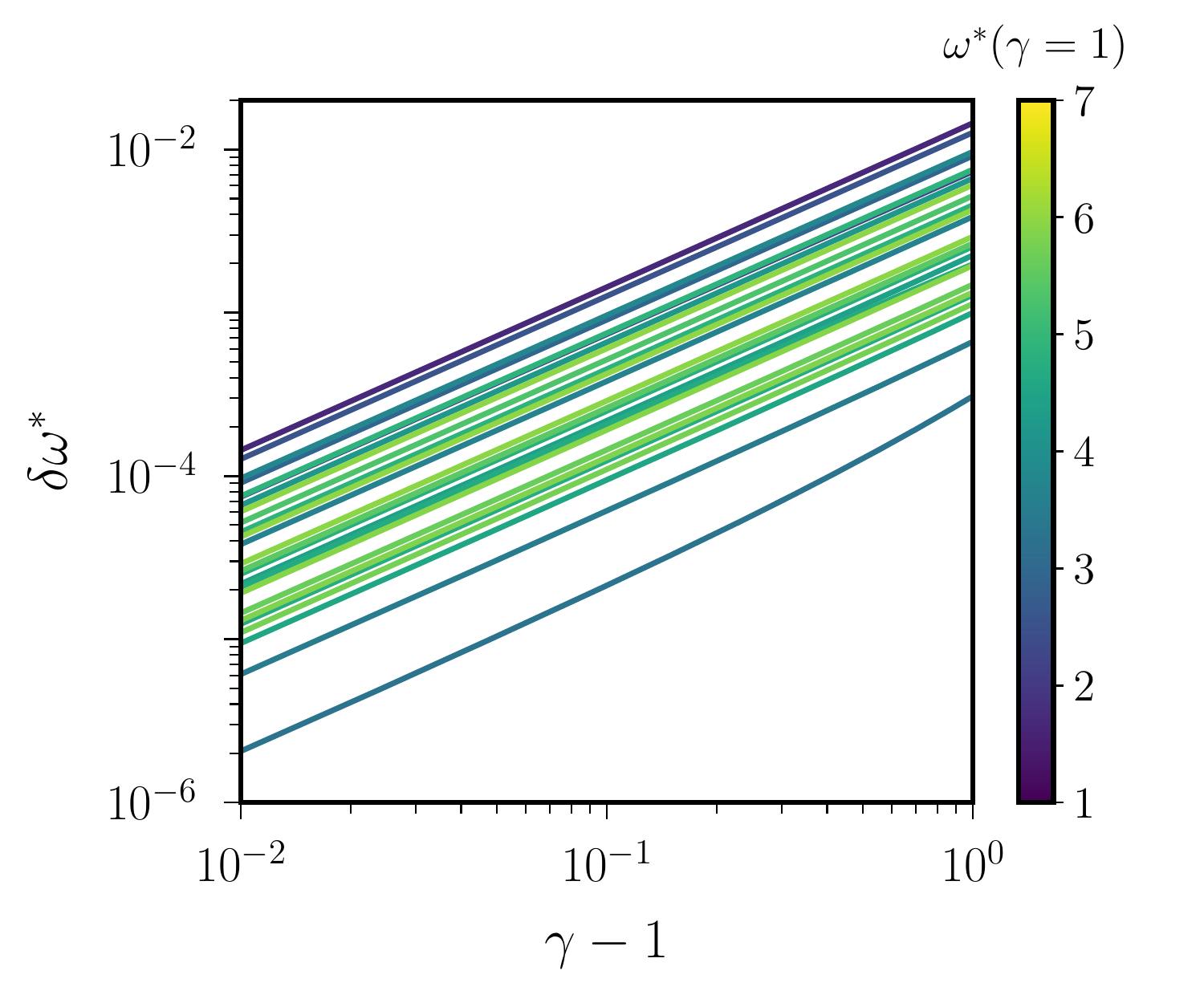}
    \caption{(Color online) Relative error $\delta \omega^* = |\omega^*-\omega^*(\gamma=1)|/|\omega^*(\gamma=1)|$, as a function of $\gamma-1$, for diffusionless acoustic modes upon isothermal states in a rotating spheroid with dimensionless polar axis $c^*=0.95$ and $M_\Omega=0.5$. Color bar indicates the angular frequency $\omega^*(\gamma=1)$ for the value $\gamma=1$.}
    \label{fig:varGamma}
\end{figure}

Finally, it is also worth considering the effects of the ambient temperature for experimental applications. 
In dimensionless variables, varying the temperature only amounts to change the values of the heat capacity ratio $\gamma \geq 1$ and of $M_\Omega$ for the isothermal regime.
The variations with $M_\Omega$ have already been considered above, so it only remains to explore the effects of $\gamma$. 
The variations of $\omega^*$ with $\gamma$ in the rapidly rotating regime are illustrated in Fig. \ref{fig:varGamma}. 
Only small departures are found from the value at $\gamma=1$, with relative errors always smaller than a percent in the range of interest $\gamma \leq 2$. 
Hence, varying $\gamma$ by changing the temperature will only have weak effects on the acoustic frequencies for usual experimental conditions with $\gamma \leq 2$.
However, the relevance of the isothermal assumption could also be questioned, since it is often difficult to keep a constant fluid temperature in experimental conditions. 
Slight variations of temperature (e.g. in the room, or due to acoustic transducers\cite{koulakis2021convective}) may indeed break the assumption of isothermal equilibrium, such that Eq. (\ref{eq:waveeqU2}) may lead to inaccurate predictions.   
Order-of-magnitude arguments show that the density variations due to hydrostatic effects would be smaller than the temperature-induced ones if\cite{koulakis2021convective} $|\rho_0 \boldsymbol{g}|~\ll~|C_s^2 \nabla \rho_0|$, that is when
\begin{subequations}
    \label{eq:koulakisinequality}
\begin{equation}
    \Delta \rho_0/\rho_0 \geq a g/C_s^2, \quad \Delta \rho_0/\rho_0 \simeq \alpha_P \Delta T,
    \tag{\theequation a,b}
\end{equation}
\end{subequations}
where $\Delta \rho_0$ is the typical variation of density across the fluid due to the temperature difference $\Delta T$ and $g$ the strength of the effective local gravity. 
If condition (\ref{eq:koulakisinequality}) is fulfilled, it could be more appropriate to use wave Eq. (\ref{eq:waveeqU1}), with possibly $\nabla P_0 = \boldsymbol{g} = \boldsymbol{0}$ to disregard the hydrostatic pressure variations due to gravity. 
If the ZoRo experiment is rotating at moderate rotation rates $\Omega/(2\pi) \simeq 15$~Hz, temperature variations of only $\Delta T \gtrsim 1$~K in air, and $\Delta T \gtrsim 5$~K in SF$_6$, could be sufficient to break the centrifugally-driven isothermal equilibrium. 
However, condition (\ref{eq:koulakisinequality}) would require $\Delta T \geq 10$~K in air and $\Delta T \geq 60$~K in SF$_6$ for $|C_s^2 \nabla \rho_0|$ to violate the (nearly) isothermal assumption for the gas-filled ZoRo experiment rotating at $\Omega/(2\pi) = 50$~Hz (with $a=0.2$~m and the centrifugal acceleration $g = g_c \sim \Omega^2 a$). 
Therefore, in the rapidly rotating regime, slight variations of temperature in ZoRo are not expected to disturb the (centrifugally-modified) acoustic resonant frequencies obtained by assuming isothermal equilibrium.

\section{Concluding remarks}
\label{sec:conclusion}
Motivated by laboratory experiments of compressible rotating flows dedicated to planetary applications\cite{triana2014helioseismology,su2020acoustic}, we have investigated in this paper how the resonant acoustic frequencies of rotating fluid-filled ellipsoids are modified by centrifugal gravity.
We have shown that, in the diffusionless theory (i.e. without attenuation), the primitive fluid-dynamic equations are exactly reducible to a single vector wave equation for the velocity perturbation upon inhomogeneous reference states in isothermal (or isentropic) equilibrium. 
Although such an approach is unconventional in acoustic modeling, it is essential to rigorously account for the effects of global rotation and buoyancy onto the acoustic perturbations. 
The acoustic problem has been numerically solved using an efficient spectral polynomial method in ellipsoids\cite{vidal2020compressible,vidal2020acoustic}, which has been validated against standard finite-element computations performed with a commercial software. 
We have shown that centrifugal gravity has non-negligible effects on the acoustic frequencies when $M_\Omega \gtrsim 0.3$ (at least for the high-frequency modes). 
Such a regime can be achieved with a rapidly rotating experiment filled with a highly compressible gas (i.e. with a speed of sound lower than that of air, such as SF$_6$ or C$_4$F$_8$). 
We have finally shown that measuring centrifugal effects would be possible in real experimental conditions, despite viscous effects and thermal diffusion.

Future applications of our work concern the on-going ZoRo experiment\cite{su2020acoustic}. 
Beyond the isothermal situation, our results for isentropic equilibrium will also be useful to predict the expected resonant acoustic frequencies upon fully turbulent convection states (since the adiabatic temperature gradient is established in compressible convection \cite{tilgner2011convection,menaut2019experimental}). 
Our results will thus be used to interpret the observed slight departures of the acoustic frequencies from our isentropic predictions in terms of convection-driven temperature anomalies upon the adiabatic gradient. 
We believe that combining experimental and theoretical works about rotating compressible flows is a promising avenue for the next generation planetary-driven convection models. 
Beyond the ZoRo experiment, our wave equation could also be considered for different boundary conditions \cite{lebovitz1989stability,vidal2019polynomial}, or for other acoustic problems with rotation (e.g. gyrometers in cylindrical geometries \citep{bruneau1986rate,ecotiere2004inertial}, which could probe the regime $M_\Omega \leq 10$ to measure very high rotation rates up to $10^{5}$ degrees per second).
Other applications of our theory could also concern (rotating) fluids that are not in isothermal nor isentropic equilibrium (e.g. with a conductive temperature profile due to internal heating \cite{koulakis2018acoustic,koulakis2021convective}), for which wave Eq. (\ref{eq:waveeqU1}) is appropriate. 
We thus hope that our results will be useful for other acoustic problems.

\section*{Acknowledgments}
We are indebted to the other members of the ZoRo team (S. Su, H.-C. Nataf, P. Cardin, M. Solazzo and Y. Do) for helpful discussions about the experiment and future prospects (rapid rotation and change of gas).
We also acknowledge the two referees for their valuable comments, which helped us to significantly improved the quality of the manuscript. 
The source code \textsc{shine} is available at \url{https://bitbucket.org/vidalje/shine/download}.
The CoolProp library\cite{bell2014pure} is available at \url{https://www.coolprop.org/}. 
This work received funding from the European Research Council (ERC) under the European Union's Horizon 2020 research and innovation programme (grant agreement No 847433, \textsc{theia} project). 

\appendix
\section{Generalized acoustic equations}
\label{appendix:waveeq}
\subsection{Mathematical formulation}
We present here a generalization of the scalar acoustic equations to account for arbitrary inhomogeneous states (i.e. possibly non-hydrostatic), but still without taking the Coriolis force into account. 
It has been initially discovered in acoustics\cite{bergmann1946wave}, and later rediscovered in astrophysics for hydrostatic reference states \cite{lignieres2009asymptotic}. 

To derive the equation, we take the time derivative of Eqs. (\ref{eq:primitiveeqn}b)-(\ref{eq:primitiveeqn}c), and replace $\lambda \boldsymbol{u}_\omega$ with Eq. (\ref{eq:primitiveeqn}a) in which the Coriolis force is neglected. 
This gives
\begin{subequations}
\label{eq:d2dt2rhoP}
\begin{align}
    \lambda^2 \rho_\omega + \nabla \boldsymbol{\cdot} (\rho_\omega \boldsymbol{g}) &= \nabla^2 P_\omega, \\
    \lambda^2 \left ( P_\omega - C_s^2 \rho_\omega \right ) &= C_s^2 \boldsymbol{S}_0 \boldsymbol{\cdot} \left [\rho_\omega \boldsymbol{g}  -\nabla P_\omega \right ],
\end{align}
\end{subequations}
where the vector $\boldsymbol{S}_0$, defined by \cite{bergmann1946wave}
\begin{equation}
\boldsymbol{S}_0 = \frac{1}{\rho_0} \left ( \nabla \rho_0 - \frac{\nabla P_0}{C_s^2} \right )  ,
\end{equation}
is related to the gradient of entropy $\nabla S_0$ in EoS (\ref{eq:backgroundS0}) such as $\nabla S_0 = (\alpha_S C_s^2) \,  \boldsymbol{S}_0$.
We then obtain from Eq. (\ref{eq:d2dt2rhoP}) the density perturbation
\begin{equation}
    \rho_\omega = \frac{\lambda^2 P_\omega + C_s^2 \boldsymbol{S}_0 \boldsymbol{\cdot} \nabla P_\omega }{C_s^2 \left ( \lambda^2 + \boldsymbol{S}_0 \boldsymbol{\cdot} \boldsymbol{g} \right )}.
    \label{eq:rhobergmann}
\end{equation}
Finally, we can substitute Eq. (\ref{eq:rhobergmann}) into Eq. (\ref{eq:d2dt2rhoP}a) to obtain the wave equation for the acoustic pressure $P_\omega$.
The pressure equation has to be supplemented by a BC, which is obtained by taking the normal component of Eq. (\ref{eq:primitiveeqn}a) on the rigid boundary $\partial V$ (where the velocity satisfies the no-penetration BC).
This gives the Robin BC
\begin{equation}
\left. \boldsymbol{n} \boldsymbol{\cdot} \left ( \nabla P_\omega - \frac{P_\omega}{C_s^2} \boldsymbol{g} \right ) \right|_{\partial V} =0.
\label{eq:BCRobin}
\end{equation}
In isentropic interiors where $\nabla P_0 = C_s^2 \, \nabla \rho_0$ and $P_\omega = C_s^2 \rho_\omega$ from EoS (\ref{eq:eosdPdR}), we obtain from (\ref{eq:d2dt2rhoP}b) the equation
\begin{equation}
    \lambda^2 P_\omega = C_s^2 \, \nabla \boldsymbol{\cdot} \left ( \nabla P_\omega - \frac{P_\omega}{C_s^2} \boldsymbol{g} \right ).
    \label{eq:waveeqP}
\end{equation}
The latter equation thus extends Eqs. (\ref{eq:helmholtz}) and (\ref{eq:waveeqPierce}) to include the effects of buoyancy for isentropic reference states (which exhibit substantial pressure variations). 
For non-isentropic reference states in hydrostatic equilibrium, we have $\nabla P_0 = \rho_0 \boldsymbol{g} \Rightarrow \nabla \rho_0 \times \boldsymbol{g} = \boldsymbol{0}$ (i.e. barotropic fluids) and the acoustic equation reduces in this case to \cite{lignieres2009asymptotic}
\begin{multline}
    \lambda^2 (\lambda^2 P_\omega + C_s^2 \boldsymbol{S}_0 \boldsymbol{\cdot} \nabla P_\omega) = C_s^2 ( \lambda^2 + N_0^2 ) \, \nabla^2 P_\omega \\
    - C_s^2 ( \lambda^2 + N_0^2 ) \, \nabla \boldsymbol{\cdot} \left ( \frac{\lambda^2 P_\omega + C_s^2 \boldsymbol{S}_0 \boldsymbol{\cdot} \nabla P_\omega}{C_s^2 (\lambda^2 + N_0^2)} \boldsymbol{g} \right ),
    \label{eq:wavePnonisentropic}
\end{multline}
where $\boldsymbol{S}_0 = (N_0^2/\boldsymbol{g}^2) \, \boldsymbol{g}$ is directly related to the square of the Brunt-V\"ais\"al\"a frequency $N_0^2 = \boldsymbol{S}_0 \boldsymbol{\cdot} \boldsymbol{g}$. 

\subsection{Range of validity of the equations}
We can now discuss the relevance of the various acoustic equations for experimental modeling. 
When the inhomogeneous reference state is isentropic, which occurs for instance in the presence of vigorous convection\cite{tilgner2011convection,menaut2019experimental} (as in planetary or stellar interiors), then Eq. (\ref{eq:waveeqP}) can be employed. 
If the reference state is not isentropic (i.e. $\nabla P_0 \neq C_s^2 \nabla \rho_0$), we can compare the magnitude of the hydrostatic pressure $|\nabla P_0| = |\rho_0 \boldsymbol{g}|$ with respect to the amplitude of $|C_s^2 \nabla \rho_0|$. 
If the fluid is in isothermal equilibrium with $\Gamma \nabla P_0 = C_s^2 \, \nabla \rho_0$ (i.e. $N_0^2 \neq 0$), then generalized acoustic Eq. (\ref{eq:wavePnonisentropic}) must be considered. 
The latter equation is however difficult to solve, because of the fourth-order polynomial degree in $\lambda$ (albeit it only involves second-order spatial derivatives for $P_\omega$). 
This is another reason showing that using vector wave Eq. (\ref{eq:waveeqU2}) is numerically advantageous, even in the non-rotating case, as it only involves a second-order time derivative.

As discussed in the main text, the isothermal assumption could become inappropriate if condition (\ref{eq:koulakisinequality}) is fulfilled. 
In such a regime, the hydrostatic pressure variations due to gravity could be disregarded to recover wave Eq. (\ref{eq:waveeqPierce}) by setting $\nabla P_0 = \boldsymbol{g} = \boldsymbol{0}$ in Eq. (\ref{eq:d2dt2rhoP}b). 
For non-rotating fluids subject to Earth's gravity, we would have $\Delta \rho_0/\rho_m \gtrsim 10^{-4}$ for dry air with $a=1$~m. 
Such small density variations can be obtained for instance with a temperature difference of $\Delta T \simeq 1$~K in air \cite{koulakis2021convective}, and so using Eq. (\ref{eq:waveeqPierce}) appears suitable in such contexts. 
However, we have shown in the main text that the Coriolis force can have significant effects on the low-frequency acoustic modes, such that Eq. (\ref{eq:waveeqPierce}) is inappropriate in the presence of global rotation. 
In the rapidly rotating regime, Eq. (\ref{eq:waveeqU2}) can be safely considered for nearly isothermal equilibrium.
Otherwise, Eq. (\ref{eq:waveeqU1}) should be considered for non-isothermal fluids in the presence of global rotation.  

\section{$\Gamma$ for isothermal equilibrium}
\label{appendix:Gamma}
We can obtain an explicit expression of the parameter $\Gamma$ in EoS (\ref{eq:eospekeris}) for fluids in isothermal equilibrium as shown below.
We rewrite Eq. (\ref{eq:primitiveeqn}b) as a function of the temperature perturbation $T_1$ upon the isothermal temperature $T_0$ in the form
\begin{equation}
    \rho_0 C_P \partial_t T_1 = \alpha_P T_0 \left[\partial_t P_1 +  \boldsymbol{u}_1 \boldsymbol{\cdot} \nabla P_0  \right],
    \label{eq:dT1}
\end{equation}
together with the linearized EoS for the perturbations
\begin{subequations}
\label{eq:eoscomsol}
\begin{equation}
    \rho_1 =  \frac{\gamma}{C_s^2} P_1 - \rho_0 \alpha_P T_1, \quad \alpha_P = \sqrt{\frac{C_P(\gamma-1)}{C_s^2 \, T_0}},
    \tag{\theequation a,b}
\end{equation}
\end{subequations}
where $C_P$ is the specific heat at constant pressure.
Then, substituting Eq. (\ref{eq:dT1}) in the time derivative of Eq. (\ref{eq:eoscomsol}a) yields an EoS in the form (\ref{eq:eosdPdR}) with 
\begin{equation}
    \Gamma = 1 + ( \alpha_P C_s)^2 T_0/C_P = \gamma \geq 1
    \label{eq:Gamma_T0}
\end{equation}
for isothermal equilibrium.
Expression (\ref{eq:Gamma_T0}) is thus valid for all (diffusionless) fluids in isothermal equilibrium. 
Note that a more straightforward derivation can be done for an ideal gas.
We have indeed $\alpha_S=1/[(1-\gamma)T_0] $ and $\alpha_P = 1/T_0$ for the ideal gas law \cite{ray1920isentropic,kouremenos1987isentropic}, which gives $\Gamma = \gamma$ according to definition (\ref{eq:Gamma}).

\section{Finite-element modeling}
\label{appendix:comsol}
To validate the spectral method presented in the main text, we have employed standard finite-element computations (performed with the commercial software \textsc{comsol}) in two cases. 
First, we have computed the diffusionless solutions of Eqs. (\ref{eq:BCRobin}) and (\ref{eq:waveeqP}), using the built-in acoustic module. 
Second, we have computed the diffusive acoustic modes for an isothermal background temperature $T_0$ and a homogeneous speed of sound $C_0$, by solving the primitive fluid-dynamic equations in the presence of viscosity and thermal diffusion. 

To do so, we include in the right-hand side of Eq. (\ref{eq:primitiveeqn}a) the viscous force $\boldsymbol{f}_v = \mu \nabla^2 \boldsymbol{u}_1 + [\mu/3 + \mu_B ] \nabla ( \nabla \boldsymbol{\cdot} \boldsymbol{u}_1)$, where $\mu$ and $\mu_B$ are respectively the dynamic shear and bulk viscosities (which are both assumed to be homogeneous in space).
We also replace Eq. (\ref{eq:primitiveeqn}b) by the linearized heat equation including thermal diffusion, which gives for isothermal equilibrium
\begin{equation}
\rho_0 C_P  \partial_t T_1 - \alpha_P T_0 \left[\partial_t P_1 +  \boldsymbol{u}_1 \boldsymbol{\cdot} \nabla P_0  \right] = k \nabla^2 T_1
\label{eq:heateq} 
\end{equation}
together with linearized EoS (\ref{eq:eoscomsol}) for the perturbations, 
where $C_P$ is assumed to be homogeneous, $k$ is the (homogeneous) thermal conductivity, and $\alpha_P$ is the (homogeneous) coefficient of thermal expansion at constant pressure defined by Eq. (\ref{eq:thermalexpcoeffs}b) and given in the general case by Eq. (\ref{eq:eoscomsol}b).
The diffusive fluid-dynamic equations are supplemented with the no-slip BC $\left . \boldsymbol{u}_1 \right |_{\partial V} =  \boldsymbol{0}$ for the velocity perturbation on the boundary, and the isothermal BC $\left . T_1 \right |_{\partial V} = 0$ for the temperature. 

The equations are discretized in \textsc{comsol} by representing the ellipsoidal geometry with an unstructured mesh made of tetrahedral finite elements.
To compute the diffusionless solutions of acoustic problem (\ref{eq:waveeqP}), we have used quintic Lagrange elements for the pressure.
For the diffusive modes, we have used cubic Lagrange elements for the pressure, and quartic Lagrange elements for the velocity and the temperature. 
For the computations of air-filled cavities, we have adopted the built-in \textsc{comsol} values for the physical parameters at the ambient temperature $T_0=20 \,^\circ$C (see also values in Table \ref{tab:thermogas}), and we have considered \cite{blackstock2001fundamentals} $\mu_B = 0.6 \mu$.

\bibliography{main}

\end{document}